\documentclass{ws-ijmpd}
\usepackage{color}

\usepackage[colorlinks,%
            citecolor=red,linkcolor=red,hypertex, %
            breaklinks=true]{hyperref}
\begin{document}

\markboth{Xian-Feng Zhao}
{The property difference between the neutron star PSR J0348+0432 and its proto neutron star}

%
\catchline{}{}{}{}{}
%

\title{The property difference between the neutron star PSR J0348+0432 and its proto neutron star
}

\author{Xian-Feng Zhao$^{1,2}$
}

\address{ $^{1}$School of Sciences, Southwest Petroleum University, Chengdu, 610500 China\\
$^{2}$School of Electronic and Electrical Engineering, Chuzhou University, Chuzhou, 239000, China
zhaopioneer.student@sina.com}



\maketitle

\begin{history}
\received{Nov 21,2016}
\revised{Day Month Year}
\comby{Managing Editor}
\end{history}

\begin{abstract}
The property difference between the neutron star PSR J0348+0432 and its proto neutron star is studied in the framework of the relativistic mean field theory considering neutrino trapping. We see that the central baryon number density of the proto neutron star PSR J0348+0432 is in the range $\rho_{c, PNS}=0.539\sim0.698$ fm$^{-3}$, which is smaller than that of the neutron star PSR J0348+0432 $\rho_{c, NS}=0.634\sim0.859$ fm$^{-3}$. Inside the neutron star PSR J0348+0432, only the neutrons, protons, $\Lambda$ and $\Xi^{-}$ produce, whereas the hyperons $\Sigma^{-}, \Sigma^{0}, \Sigma^{+}$ and $\Xi^{0}$ all do not appear. But in the proto neutron star PSR J0348+0432, hyperons $\Sigma^{-}$, $\Sigma^{0}$, $\Sigma^{+}$ and $\Xi^{0}$ all will produce, though their relative particle number density are still very small, no more than 2\%. This shows that higher temperature will be advantageous to the hyperon production.
\end{abstract}


\section{Introduction}

In the supernova, core implosion will take 0.5$\sim$1.0 s and within a few milliseconds the lepton-rich core will settle into hydrostatic equilibrium. Thus, a proto neutron star (PNS) produces. Only after tens of seconds, it will change into a neutron star (NS). For a PNS with $M$=1.4 M$_{\odot}$, within $t$=0$\sim$20 s, its temperature $T$ changes from 30 MeV to 5 MeV~\cite{Burrows86}.

The phase diagram at sub-saturation densities is strongly affected by the electromagnetic interaction, though it is almost independent of the electric charge at supra-saturation density~\cite{Gulminelli13}. Although the PNS has strong magnetic field, but its effect on the dynamics is found to be mild~\cite{Sawai13}. The symmetry energy plays a dramatic role in determining the
structure of NSs and the evolution of core-collapsing supernovae~\cite{Raduta14}. And at the end of the evolution, an isolated NS has a certain maximum value of mass, which it can not reach~\cite{Camelio16}.

Within the finite-temperature Brueckner-Bethe-Goldstone many-body theory, the research results show that neutrino trapping will shift the appearance of hyperons to larger baryon density and will stiffen considerably the equation of state~\cite{Burgio07,Nicotra06}.

Recent years, two massive NSs have been discovered one after another. In 2010, the NS PSR J1614-2230 was observed by Demorest et al~\cite{Demorest10} and the other more massive NS PSR J0348+0432, whose mass is $M$=2.01$\pm0.04$ M$_{\odot}$ and is almost the largest one by far, was also found by Antoniadis et al in 2013~\cite{Antoniadis13}.
The research results show that a non-linear relativistic mean field (RMF) model, which is consistent with up-to-date semi-empirical nuclear and hypernuclear data, can allows for neutron stars with hyperon cores and $M >$ 2 M$_{\odot}$. This model involves hidden-strangeness scalar and vector mesons, coupled only to hyperons, and quartic terms involving vector meson fields~\cite{Bednarek12}.

The mass of the massive NS/PNS is very large and this will no doubt limit the NS/PNS's properties, such as the field strength of meosons, chemical potentials of baryons, particle distribution and the NS/PNS's mass and radius.

For the massive PNSs, their properties are very important for the understanding of the stellar evolution. The PNS has higher temperature compared to the NS and therefore its properties must be different from those of its corresponding NS. About this case, we have greater interest.

For finite temperatures above ~1 MeV, the neutrino mean free path in neutron star
matter is less than the radius of the star so that neutrinos are trapped and have to be taken into
account in the beta-equilibrium. So the calculations for the proto-neutron star case should be done
for conserved lepton number~\cite{Prakash97}.

In this paper, we use the RMF theory to study the property difference between the NS PSR J0348+0432 and its PNS with neutrino trapping being considered.

\section{The RMF theory}
Under the RMF approximation, the Lagrangian density of hadron matter considering the mesons $f_{0}(975)$ (denoted as $\sigma^{*}$) and $\phi(1020)$ (denoted as $\phi$) reads as follows~\cite{Schaffner94,Glendenning97}
\begin{eqnarray}
\mathcal{L}&=&
\sum_{B}\overline{\Psi}_{B}(i\gamma_{\mu}\partial^{\mu}-{m}_{B}+g_{\sigma B}\sigma+g_{\sigma^{*}B}\sigma^{*}-g_{\omega B}\gamma^{0}\omega-g_{\phi B}\gamma^{0}\phi-g_{\rho B}\gamma^{0}\tau_{3}\rho)\Psi_{B}
\nonumber\\
&&-\frac{1}{2}m_{\sigma}^{2}\sigma^{2}-\frac{1}{3}g_{2}\sigma^{3}-\frac{1}{4}g_{3}\sigma^{4}
+\frac{1}{2}m_{\omega}^{2}\omega^{2}+\frac{1}{2}m_{\rho}^{2}\rho^{2}-\frac{1}{2}m_{\sigma^{*}}^{2}\sigma^{*2}
+\frac{1}{2}m_{\phi}^{2}\phi^{2}
\nonumber\\
&&+\sum_{\lambda=e,\mu}\overline{\Psi}_{\lambda}\left(i\gamma_{\mu}\partial^{\mu}
-m_{\lambda}\right)\Psi_{\lambda}
.\
\end{eqnarray}

For the PNS matter considering neutrino trapping, the partition function of baryon reads as

\begin{eqnarray}
lnZ_{B}=\frac{V}{T}\langle\mathcal{L}\rangle+\sum_{B}\frac{2J_{B}+1}{2\pi^{2}}\int_{0}^{\infty}k^{2}dk
\left\{ln\left[1+e^{-(\varepsilon_{B}(k)-\mu_{B})/T}\right]\right\}
\end{eqnarray}

From above we can obtain the total baryon number density, the energy density and the pressure as follows~\cite{GlendenningPlb87,Glendenningnpa87,Reddy98}

\begin{eqnarray}
\rho=\sum_{B}\frac{2J_{B}+1}{2\pi^{2}}b_{B}\int_{0}^{\infty}k^{2}n_{B}(k)dk,
\end{eqnarray}

\begin{eqnarray}
\varepsilon&=&\frac{1}{2}
m_{\sigma}^{2}\sigma^{2}+\frac{1}{2}
m_{\sigma^{*}}^{2}\sigma^{*2}+\frac{1}{3}g_{2}\sigma^{3}+\frac{1}{4}g_{3}\sigma^{4}
+\frac{1}{2}m_{\omega}^{2}\omega_{0}^{2}+\frac{1}{2}m_{\phi}^{2}\phi^{2}
+\frac{1}{2}m_{\rho}^{2}\rho_{03}^{2}
\nonumber\\
&&+\sum_{B}\frac{2J_{B}+1}{2\pi^{2}}\int_{0}^{\infty}\kappa^{2}n_{B}(k)\mathrm d\kappa\sqrt{\kappa^{2}+m_{B}^{*2}},
\\
p&=&-\frac{1}{2}m_{\sigma}^{2}\sigma^{2}-\frac{1}{2}m_{\sigma^{*}}^{2}\sigma^{*2}-\frac{1}{3}g_{2}\sigma^{3}
-\frac{1}{4}g_{3}\sigma^{4}
+\frac{1}{2}m_{\omega}^{2}\omega_{0}^{2}+\frac{1}{2}m_{\phi}^{2}\phi^{2}
+\frac{1}{2}m_{\rho}^{2}\rho_{03}^{2}
\nonumber\\
&&+\frac{1}{3}\sum_{B}\frac{2J_{B}+1}{2\pi^{2}}\int_{0}^{\infty}\frac{\kappa^{4}}{\sqrt{\kappa^{2}
+m_{B}^{*2}}}n_{B}(k)d\kappa,
\end{eqnarray}
where, $n_{B}(k)$ is the Fermi-Dirac partition function of baryon
\begin{eqnarray}
n_{B}(k)=\frac{1}{1+exp\left[\left(\varepsilon_{B}(k)-\mu_{B}\right)/T\right]}.
\end{eqnarray}

For the leptons, their interactions can be ignored at finite temperature and therefore their partition function
read as

\begin{eqnarray}
lnZ_{L}&=&\frac{V}{T}\sum_{i}\frac{\mu_{i}^{4}}{24\pi^{2}}\left[1+2\left(\frac{\pi T}{\mu_{i}}\right)^{2}+\frac{7}{15}\left(\frac{\pi T}{\mu_{i}}\right)^{4}\right]
\nonumber\\
&&+V\sum_{\lambda}\frac{1}{\pi^{2}}\int_{0}^{\infty}k^{2}dk
\left\{ln\left[1+e^{-(\varepsilon_{\lambda}(k)-\mu_{\lambda})/T}\right]\right\},
\end{eqnarray}
the first line represents the contribution of massless neutrinos and the second line the contribution of electrons
and $\mu$s.

From above we can obtain the lepton number density and the contribution to the energy density and the pressure

\begin{eqnarray}
\rho_{l}=\frac{1}{\pi^{2}}\int_{0}^{\infty}k^{2}n_{l}(k)dk,
\\
\rho_{\nu}=\frac{\pi^{2}T^{2}\mu_{\nu}+\mu_{\nu}^{3}}{6\pi^{2}},
\end{eqnarray}

\begin{eqnarray}
\varepsilon=\sum_{l}\frac{1}{\pi^{2}}\int_{0}^{\infty}\kappa^{2}n_{l}(k)\mathrm d\kappa\sqrt{\kappa^{2}+m_{l}^{2}}=\sum_{\nu}\left(\frac{7\pi^{2}T^{4}}{120}+\frac{T^{2}\mu^{2}_{\nu}}{4}+
\frac{\mu_{\nu}^{4}}{8\pi^{2}}\right),
\end{eqnarray}

\begin{eqnarray}
p=\frac{1}{3}\sum_{l}\frac{1}{\pi^{2}}\int_{0}^{\infty}\frac{\kappa^{4}}{\sqrt{\kappa^{2}
+m_{l}^{2}}}n_{l}(k)d\kappa=\sum_{\nu}\frac{1}{360}\left(7\pi^{2}T^{4}+30T^{2}\mu^{2}_{\nu}+
\frac{15\mu_{\nu}^{4}}{\pi^{2}}\right).
\end{eqnarray}

The chemical potentials of baryons can be written as

\begin{eqnarray}
\mu_{i}=\mu_{n}-q_{i}\left(\mu_{e}-\mu_{\nu e}\right).
\end{eqnarray}

To describe the neutrino trapping, we can define the content of lepton in PNS as

\begin{eqnarray}
Y_{l \nu}=Y_{l}+Y_{\nu l}=\frac{\rho_{l}+\rho_{\nu l}}{\rho}.
\end{eqnarray}

For the NS, the RMF theory can be seen in Ref [\cite{Glendenning97}].

We use {\color{blue}the Tolman-Oppenheimer-Volkoff (TOV) equation} to obtain the mass and the radius of a NS/PNS{\color{blue}~\cite{Tolman39,Oppenheimer39}}
\begin{eqnarray}
\frac{\mathrm dp}{\mathrm dr}&=&-\frac{\left(p+\varepsilon\right)\left(M+4\pi r^{3}p\right)}{r \left(r-2M \right)}
,\\\
M&=&4\pi\int_{0}^{{\color{blue}R}}\varepsilon r^{2}\mathrm dr
.\
\end{eqnarray}
{\color{blue}Here, the upper integration limit $R$ is the radius of the star which contains the accumulated total mass $M$.}

\section{The parameters}
In this work, the nucleon coupling constant is chosen as the GL85 set~\cite{Glendenning85}: the saturation density $\rho_{0}$=0.145 fm$^{-3}$, binding energy B/A=15.95 MeV, a compression modulus $K=285$ MeV, charge symmetry coefficient $a_{sym}$=36.8 MeV and the effective mass $m^{*}/m$=0.77.

We define the ratios of hyperon coupling constants to nucleon coupling constants as follows: $x_{\sigma h}=\frac{g_{\sigma h}}{g_{\sigma}}$, $x_{\omega h}=\frac{g_{\omega h}}{g_{\omega}}$, $x_{\rho h}=\frac{g_{\rho h}}{g_{\rho}}
$, with $h$ denoting hyperons $\Lambda, \Sigma$ and $\Xi$.

According to SU(6) symmetry, $x_{\rho \Lambda}=0, x_{\rho \Sigma}=2, x_{\rho \Xi}=1$ are selected~\cite{Schaff96}. The experimental data show that the hyperon well depth are $U_{\Lambda}^{(N)}=-30$ MeV~\cite{Batt97}, $ U_{\Sigma}^{(N)}=10\sim40$ MeV~\cite{{Kohno06},{Harada05},{Harada06},{Fried07}} and $U_{\Xi}^{(N)}=-18$ MeV~\cite{Schaff00}, respectively. Therefore, we choose $U_{\Lambda}^{(N)}=-30$ MeV, $ U_{\Sigma}^{(N)}$=+30 MeV and $U_{\Xi}^{(N)}=-18$ MeV in this work.

The calculating results show that the ratio of hyperon coupling constant to nucleon coupling constant is in the range of $\sim$ 1/3 to 1~\cite{Glen91}. So in this work we choose $x_{\sigma \Lambda}$=0.4, 0.5, 0.6, 0.7, 0.8, 0.9 at first and the hyperon coupling constants $x_{\omega \Lambda}$ can be obtained by considering the restriction of the hyperon well depth~\cite{Glen97}

\begin{eqnarray}
U_{h}^{(N)}=m_{n}\left(\frac{m_{n}^{*}}{m_{n}}-1\right)x_{\sigma h}+\left(\frac{g_{\omega}}{m_{\omega}}\right)^{2}\rho_{0}x_{\omega h}
.\
\end{eqnarray}

Using the same method, $x_{\sigma \Sigma}$, $x_{\omega \Sigma}$, $x_{\sigma \Xi}$ and $x_{\omega \Xi}$ can be obtained (see Table~\ref{tab1}).

The coupling constants of the mesons $f_{0}(975)$ and $\phi(1020)$ are $g_{\phi \Xi}=2g_{\phi \Lambda}=-2\sqrt{2}g_{\omega}/3$ and $g_{f_{0} \Lambda}/g_{\sigma}=g_{f_{0} \Sigma}/g_{\sigma}=0.69$, $g_{f_{0} \Xi}/g_{\sigma}=1.25$, respectively~\cite{Schaffner94}.

\begin{table}[!htbp]
\tbl{The hyperon coupling constants fitting to the experimental data of the well depth, which are $U_{\Lambda}^{(N)}=-30$ MeV, $U_{\Sigma}^{(N)}=+30$ MeV and $U_{\Xi}^{(N)}=-18$ MeV, respectively. }
{\begin{tabular}{@{}cccccc@{}} \toprule
$x_{\sigma \Lambda}$ &$x_{\omega \Lambda}$&$x_{\sigma \Sigma}$&$x_{\omega \Sigma}$&$x_{\sigma \Xi}$&$x_{\omega \Xi}$ \\
\hline
0.4               &0.3681 &0.4 &0.7597&0.4               &0.4464 \\
0.5               &0.5090 &0.5 &0.9007&0.5               &0.5874 \\
0.6               &0.6500 &    &      &0.6               &0.7284 \\
0.7               &0.7910 &    &      &0.7               &0.8693 \\
0.8               &0.9320 &    &      &                  & \\
\hline
\end{tabular} \label{tab1}}
\end{table}

In this work, the entropy per baryon is chosen as S=2. The content of lepton in PNS is chosen as $Y_{l\mu}$=0 and
$Y_{le}$=0.1, respectively.

From Table~\ref{tab1}, we can combine into 40 sets of parameters (see Table~\ref{tab2}).

\begin{table}[!htbp]
\tbl{The 44 sets of hyperon coupling constants used in this work and the corresponding NS/PNS's maximum mass claculated. The unit of the mass $M$ is the solar mass M$_{\odot}$. }
{\begin{tabular}{@{}ccccccccc@{}} \toprule
NO.&$x_{\sigma \Lambda}$ &$x_{\omega \Lambda}$&$x_{\sigma \Sigma}$&$x_{\omega \Sigma}$&$x_{\sigma \Xi}$     &$x_{\omega \Xi}$&$M_{max,PNS}$&$M_{max,NS}$\\
\hline
01&0.4     &0.3681 &0.4             &0.7597            &0.4  &0.4464&1.5427&1.4300\\
02&0.4     &0.3681 &0.4             &0.7597            &0.5  &0.5874&1.5577&1.4303\\
03&0.4     &0.3681 &0.4             &0.7597            &0.6  &0.7284&1.5690&1.4303\\
04&0.4     &0.3681 &0.4             &0.7597            &0.7  &0.8693&1.5771&1.4303\\
05&0.4     &0.3681 &0.5             &0.9007            &0.4  &0.4464&1.5434&1.4300\\
06&0.4     &0.3681 &0.5             &0.9007            &0.5  &0.5874&1.5586&1.4303\\
07&0.4     &0.3681 &0.5             &0.9007            &0.6  &0.7284&1.5700&1.4303\\
08&0.4     &0.3681 &0.5             &0.9007            &0.7  &0.8693&1.5783&1.4303\\
09&0.5     &0.5090 &0.4             &0.7597            &0.4  &0.4464&1.6420&1.5576\\
10&0.5     &0.5090 &0.4             &0.7597            &0.5  &0.5874&1.6691&1.5705\\
11&0.5     &0.5090 &0.4             &0.7597            &0.6  &0.7284&1.6888&1.5746\\
12&0.5     &0.5090 &0.4             &0.7597            &0.7  &0.8693&1.7021&1.5747\\
13&0.5     &0.5090 &0.5             &0.9007            &0.4  &0.4464&1.6430&1.5576\\
14&0.5     &0.5090 &0.5             &0.9007            &0.5  &0.5874&1.6705&1.5705\\
15&0.5     &0.5090 &0.5             &0.9007            &0.6  &0.7284&1.6907&1.5746\\
16&0.5     &0.5090 &0.5             &0.9007            &0.7  &0.8693&1.7045&1.5747\\
17&0.6     &0.6500 &0.4             &0.7597            &0.4  &0.4464&1.7368&1.6780\\
18&0.6     &0.6500 &0.4             &0.7597            &0.5  &0.5874&1.7838&1.7189\\
19&0.6     &0.6500 &0.4             &0.7597            &0.6  &0.7284&1.8182&1.7460\\
20&0.6     &0.6500 &0.4             &0.7597            &0.7  &0.8693&1.8403&1.7570\\
21&0.6     &0.6500 &0.5             &0.9007            &0.4  &0.4464&1.7382&1.6780\\
22&0.6     &0.6500 &0.5             &0.9007            &0.5  &0.5874&1.7860&1.7189\\
23&0.6     &0.6500 &0.5             &0.9007            &0.6  &0.7284&1.8215&1.7460\\
24&0.6     &0.6500 &0.5             &0.9007            &0.7  &0.8693&1.8447&1.7570\\
25&0.7     &0.7910 &0.4             &0.7597            &0.4  &0.4464&1.8011&1.7590\\
26&0.7     &0.7910 &0.4             &0.7597            &0.5  &0.5874&1.8683&1.8318\\
27&0.7     &0.7910 &0.4             &0.7597            &0.6  &0.7284&1.9217&1.8904\\
28&0.7     &0.7910 &0.4             &0.7597            &0.7  &0.8693&1.9579&1.9249\\
29&0.7     &0.7910 &0.5             &0.9007            &0.4  &0.4464&1.8030&1.7590\\
30&0.7     &0.7910 &0.5             &0.9007            &0.5  &0.5874&1.8714&1.8318\\
31&0.7     &0.7910 &0.5             &0.9007            &0.6  &0.7284&1.9268&1.8904\\
32&0.7     &0.7910 &0.5             &0.9007            &0.7  &0.8693&1.9652&1.9249\\
33&0.8     &0.9320 &0.4             &0.7597            &0.4  &0.4464&1.8329&1.7911\\
34&0.8     &0.9320 &0.4             &0.7597            &0.5  &0.5874&1.9114&1.8845\\
35&0.8     &0.9320 &0.4             &0.7597            &0.6  &0.7284&1.9791&1.9690\\
36&0.8     &0.9320 &0.4             &0.7597            &0.7  &0.8693&2.0297&2.0275\\
37&0.8     &0.9320 &0.5             &0.9007            &0.4  &0.4464&1.8354&1.7911\\
38&0.8     &0.9320 &0.5             &0.9007            &0.5  &0.5874&1.9156&1.8845\\
39&0.8     &0.9320 &0.5             &0.9007            &0.6  &0.7284&1.9863&1.9694\\
40&0.8     &0.9320 &0.5             &0.9007            &0.7  &0.8693&2.0407&2.0300\\
\hline
41&0.8     &0.9320 &0.5             &0.9007            &0.71 &0.8834&2.0452&2.0343\\
42&0.8     &0.9320 &0.5             &0.9007            &0.73 &0.9116&2.0536&2.0419\\
43&0.8     &0.9320 &0.5             &0.9007            &0.75 &0.9398&2.0612&2.0483\\
44&0.8     &0.9320 &0.5             &0.9007            &0.77 &0.9680&2.0682&2.0535\\
\hline
\end{tabular} \label{tab2}}
\end{table}

\begin{figure}[!htp]
\centering{}\includegraphics[width=4.5in]{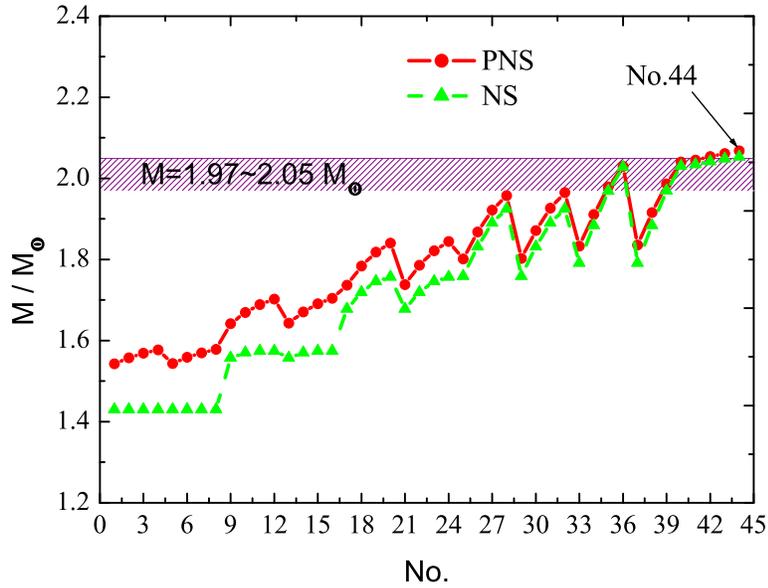}\caption{The mass as a function of the parameters numbers.}
\label{fig1}
\end{figure}

For each set of parameters in Table~\ref{tab2}, we calculate the mass of the NS/PNS (see Fig.~\ref{fig1}).

We find that the maximum mass of the NS calculated by parameters 01 to 40 are all less than 2.05 M$_{\odot}$. Then we further choose $x_{\sigma \Xi}$=0.71, 0.73, 0.75, 0.77, respectively. Thus, we obtain parameters 41 to 44 (see Table~\ref{tab2}), by which the maximum masses obtained are also listed in Table~\ref{tab2}. We see that parameter 44 can give the maximum mass greater than 2.05 M$_{\odot}$. Therefore, we use parameter 44 to describe the properties of the NS/PNS PSR J0348+0432.

\section{The field strength of mesons $\sigma$, $\omega$ and $\rho$}
Our calculations show that the central baryon number density of the NS PSR J0348+0432 is $\rho_{Bc,NS}$=0.634$\sim$0.859 fm$^{-3}$ and that of the PNS PSR J0348+0432 is $\rho_{Bc,PNS}$=0.539$\sim$0.698 fm$^{-3}$ (see Table~\ref{tab3}).

\begin{table}[!htbp]
\tbl{The results of the NS/PNS PSR J0348+0432 calculated in this work.
The unit of the following physical quantities are: $M$$\sim$M$_{\odot}$; $\rho_{c}$$\sim$fm$^{-3}$; $\sigma_{0}$, $\omega_{0}$, $\rho_{03}$$\sim$fm$^{-1}$; $\mu_{nc}$, $\mu_{ec}$$\sim$fm$^{-1}$, respectively.}
{\begin{tabular}{@{}ccc@{}} \toprule
                               &NS PSR J0348+0432           &PNS PSR J0348+0432  \\
\hline
$M$                            &1.97$\sim$2.05               &1.97$\sim$2.05 \\
\hline
$\rho_{Bc}$                 &0.634$\sim$0.859&0.539$\sim$0.698 \\
\hline
$\sigma_{c}$                 &0.321$\sim$0.366&0.290$\sim$0.326 \\
$\omega_{c}$                 &0.367$\sim$0.492&0.311$\sim$0.401 \\
$\rho_{c}$                   &0.087$\sim$0.091&0.079$\sim$0.079 \\
\hline
$\mu_{nc}$                    &6.957$\sim$7.906&6.395$\sim$7.046 \\
$\mu_{ec}$                    &1.314$\sim$1.294&1.057$\sim$1.142 \\
\hline
$\rho_{nc}/\rho$      &64.3$\sim$48.3\%&60.9$\sim$52.1\% \\
$\rho_{pc}/\rho$      &21.3$\sim$21.4\%&17.6$\sim$18.7\% \\
$\rho_{\Lambda c}/\rho$&14.4$\sim$23.8\%&13.5$\sim$17.6\% \\
$\rho_{\Sigma^{-}c}/\rho$&0$\sim$0\%   &0.06\%$\sim$0.1\% \\
$\rho_{\Sigma^{0}c}/\rho$&0$\sim$0\%    &0.2\%$\sim$0.4\% \\
$\rho_{\Sigma^{+}c}/\rho$&0$\sim$0\%    &0.8$\sim$1.4\% \\
$\rho_{\Xi^{-}c}/\rho$&0$\sim$6.5\%     &6.3$\sim$8.4\% \\
$\rho_{\Xi^{0}c}/\rho$&0$\sim$0\%       &0.7$\sim$1.2\% \\
\hline
$\rho_{ec}/\rho$&12.1$\sim$8.5\%        &10.3$\sim$10.1\% \\
$\rho_{\mu c}/\rho$&9.2$\sim$6.4\%      &1.8$\sim$1.5\% \\
\hline
\end{tabular} \label{tab3}}
\end{table}

\begin{figure}[!htp]
\centering{}\includegraphics[width=4.5in]{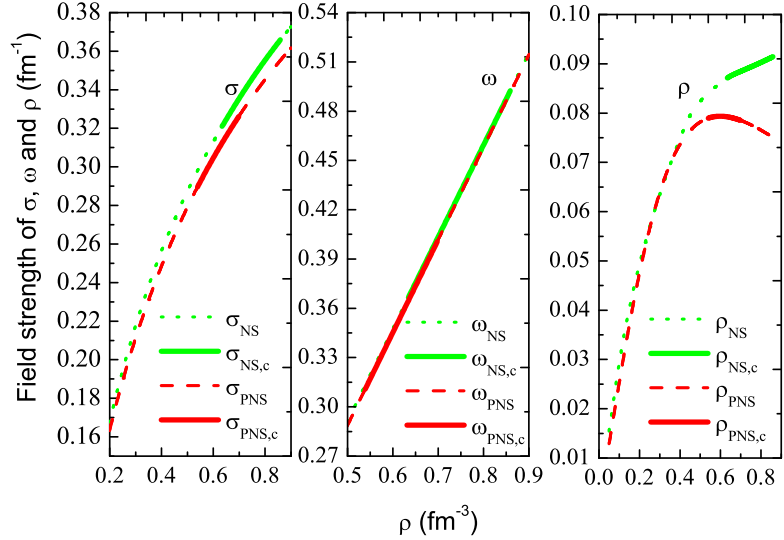}\caption{The field strength of mesons $\sigma$, $\omega$ and $\rho$ as a function of the baryon number density.}
\label{fig2}
\end{figure}

The field strengths of mesons $\sigma$, $\omega$ and $\rho$ as a function of the baryon number density are shown in Fig.~\ref{fig2}. The think lines show the central field strengths.

We see that the field strengths of mesons $\sigma$ in the PNS PSR J0348+0432 are less than those in the NS PSR J0348+0432. At the center, the field strength of mesons $\sigma$ of the PNS PSR J0348+0432 is in the range $\sigma_{PNSc}$=0.290$\sim$0.326 fm$^{-1}$, which is slightly smaller than that of the NS PSR J0348+0432 (in the range $\sigma_{NSc}$=0.321$\sim$0.366 fm$^{-1}$).

The field strength of mesons $\omega$ in the PNS PSR J0348+0432 and that in the NS PSR J0348+0432 make little difference corresponding to a same baryon number density $\rho$. But the central field strength of mesons $\omega$ of the NS PSR J0348+0432 is in the range $\omega_{NSc}$=0.367$\sim$0.492 fm$^{-1}$, while that of the PNS PSR J0348+0432 decreases to the range $\omega_{PNSc}$=0.311$\sim$0.401 fm$^{-1}$.

We also see that there are little differences between the field strength of meson $\rho$ of the PNS PSR J0348+0432 and that of the NS PSR J0348+0432 corresponding to a same baryon number density $\rho$. In the center , the field strength of mesons $\rho$ of the PNS PSR J0348+0432 is $\rho_{PNSc}$=0.079 fm$^{-1}$, while that of the NS PSR J0348+0432 is in the range $\rho_{NSc}$=0.087$\sim$0.091 fm$^{-1}$.

From the above we see that the field strengths of mesons $\sigma, \omega$ and $\rho$ of the PNS PSR J0348+0432 and those in the NS PSR J0348+0432 all make little differences corresponding to a same baryon number density $\rho$. But the value range of the central field strength of mesons $\sigma, \omega$ and $\rho$ in the PNS PSR J0348+0432 is smaller than those in the NS PSR J0348+0432.

The central field strengths of mesons $\sigma, \omega$ and $\rho$ also can be seen in Table~\ref{tab3}.

\section{The chemical potentials of neutrons and electrons}
\begin{figure}[!htp]
\centering{}\includegraphics[width=4.5in]{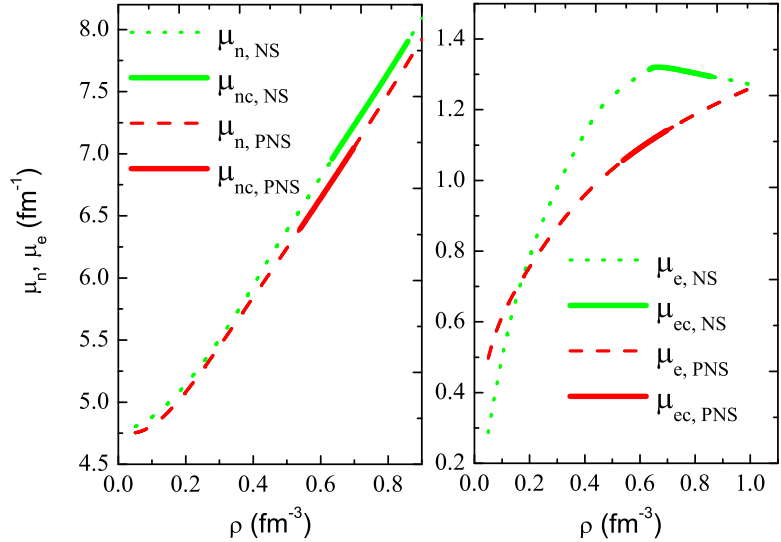}\caption{The chemical potentials of neutrons $\mu_{n}$ and electrons $\mu_{e}$¡¡as a function of baryon number density.}
\label{fig3}
\end{figure}

The chemical potentials of neutrons and electrons as a function of baryon number density are shown in Fig.~\ref{fig3}. The think lines show the central chemical potentials. We see the chemical potentials of neutrons of the PNS PSR J0348+0432 are less than those of the NS PSR J0348+0432 corresponding to a same baryon number density $\rho$. As $\rho<0.18$ fm$^{-3}$ the chemical potentials of electrons of the PNS PSR J0348+0432 are greater than those of the NS PSR J0348+0432 but are less than the latter as $\rho>0.18$ fm$^{-3}$.

The central chemical potential of neutrons of the PNS PSR J0348+0432 is in the range $\mu_{nc, PNS}$=6.395$\sim$7.046 fm$^{-1}$, while that of the NS PSR J0348+0432 is in the range $\mu_{nc, NS}$=6.957$\sim$7.906 fm$^{-1}$. The central chemical potential of electrons of the PNS PSR J0348+0432 is in the range $\mu_{ec,PNS}$=1.057$\sim$1.142 fm$^{-1}$, while that of the NS PSR J0348+0432 is in the range $\mu_{ec,NS}$=1.314$\sim$1.294 fm$^{-1}$.

\section{The relative particle number density of the baryons in the NS/PNS PSR J0348+0432}
\subsection{n, p, $\Lambda$}
\begin{figure}[!htp]
\centering{}\includegraphics[width=4.5in]{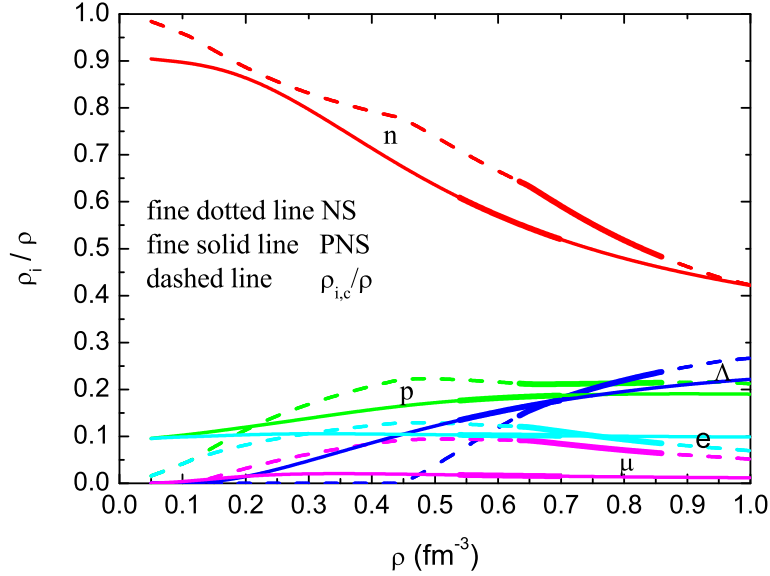}\caption{The relative particle number density of n, p, $\Lambda$, e and $\mu$ as a function of baryon number density.}
\label{fig4}
\end{figure}

Figure~\ref{fig4} gives the relative particle number density $\rho_{i}/\rho$ of neutrons, protons, $\Lambda$, electrons and $\mu$s as a function of baryon number density $\rho$.

We see the relative particle number density of neutrons $\rho_{n}/\rho$ in the PNS PSR J0348+0432 is less than those in the NS PSR J0348+0432 corresponding to a same baryon number density $\rho$. The relative particle number density of protons $\rho_{p}/\rho$ in the PNS PSR J0348+0432 is greater than those in the NS PSR J0348+0432 as $\rho<0.2083$ fm$^{-3}$ but is less than the latter as $\rho>0.2083$ fm$^{-3}$. The relative particle number density of $\Lambda$ $\rho_{\Lambda}/\rho$ in the PNS PSR J0348+0432 is greater than that in the NS PSR J0348+0432.

The central relative particle number density of neutrons of the PNS PSR J0348+0432 is in the range $\rho_{nc,PNS}/\rho$=60.9$\sim$52.1\%, while that of the NS PSR J0348+0432 is in the range $\rho_{nc,NS}/\rho$=64.3$\sim$48.3\%.

The relative particle number density of protons at the center of the PNS PSR J0348+0432 is in the range $\rho_{pc,PNS}/\rho$=17.6$\sim$18.7\%, while that of the NS PSR J0348+0432 is in the range $\rho_{pc,NS}/\rho$=21.3$\sim$21.4\%.

The relative particle number density of $\Lambda$s at the center of the PNS PSR J0348+0432 is in the range $\rho_{\Lambda c,PNS}/\rho$=13.5$\sim$17.6\%, while that of the NS PSR J0348+0432 is in the range $\rho_{\Lambda c,NS}/\rho$=14.4$\sim$23.8\%.

From the above we see that the central relative number density of neutrons and protons in the PNS PSR J0348+0432 are all less than those in the NS PSR J0348+0432.

\subsection{$\Sigma^{-}$, $\Sigma^{0}$, $\Sigma^{+}$}
In the NS PSR J0348+0432, the hyperons $\Sigma^{-}, \Sigma^{0}$ and $\Sigma^{+}$ all do not produce. But in the PNS PSR J0348+0432, they will all produce, though their relative particle number density are still very small, only less than 2\%.

This means that although the positive well depth $U_{\Sigma}^{(N)}$ restricts the production of the $\Sigma$~\cite{zhaoprc15,zhaoprc12}, but the higher temperature will be advantageous to its production.

\begin{figure}[!htp]
\centering{}\includegraphics[width=4.5in]{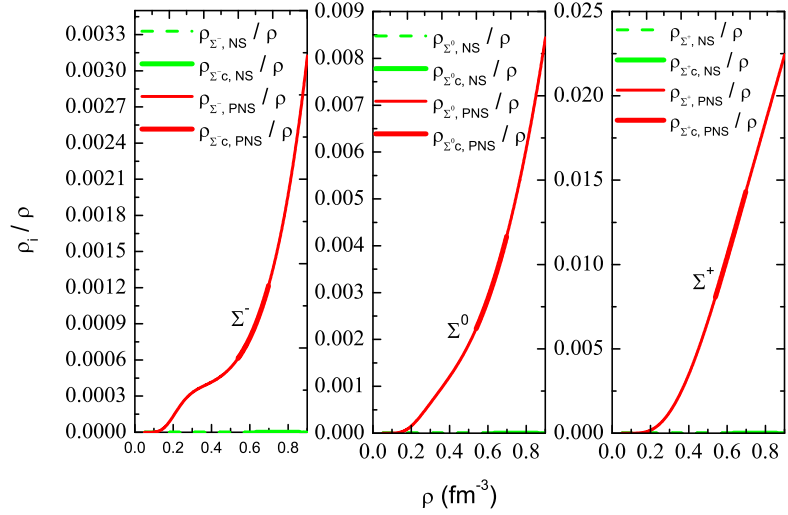}\caption{The relative particle number density of $\Sigma$ as a function of baryon number density.}
\label{fig5}
\end{figure}

The relative particle number density of hyperon $\Sigma$ as a function of baryon number density $\rho$ is shown in Fig.~\ref{fig5}.

\subsection{$\Xi^{-}$, $\Xi^{0}$}
\begin{figure}[!htp]
\centering{}\includegraphics[width=4.5in]{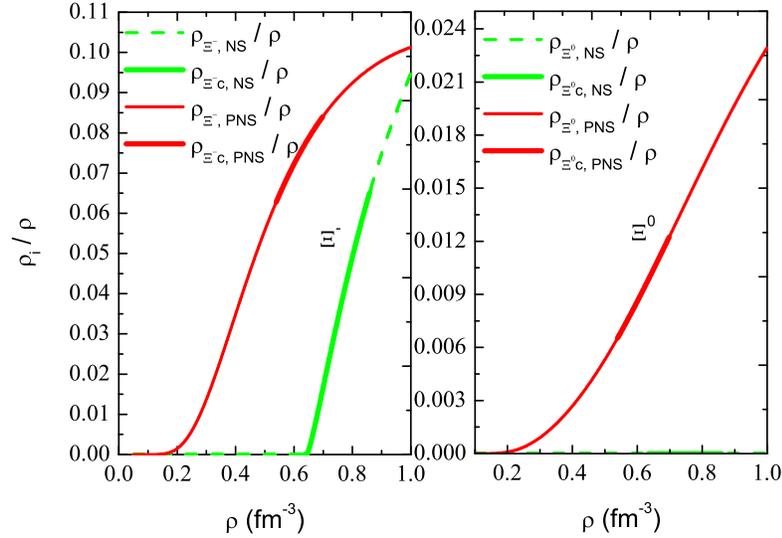}\caption{The relative particle number density of $\Xi^{-}$ and $\Xi^{0}$ as a function of baryon number density.}
\label{fig6}
\end{figure}

The relative particle number density of $\Xi^{-}$ and $\Xi^{0}$ as a function of baryon number density are shown in Fig~\ref{fig6}.

We see that hyperons $\Xi^{-}$ and $\Xi^{0}$ all will produce in the PNS PSR J0348+0432 whereas in the NS PSR J0348+0432 only $\Xi^{-}$ appears. The central relative particle number density of $\Xi^{-}$s in the NS PSR J0348+0432 is in the range $\rho_{\Xi^{-}c, NS}/\rho$=0$\sim$6.5\%, while that in the PNS PSR J0348+0432 is in the larger range $\rho_{\Xi^{-}c,PNS}/\rho$=6.3$\sim$8.4\%. Higher temperature will be advantageous to the production of hyperon $\Xi^{-}$.

In the NS PSR J0348+0432, the hyperons $\Xi^{0}$ do not produce. But in the PNS PSR J0348+0432 it produces, though the relative particle number density is still very small. In the center, the relative particle number density is only $\rho_{\Xi^{0}c,PNS}/\rho$=0.7\%$\sim$1.2\%, namely less than 2\%.

\section{Effect of temperature on the relative particle number density of the baryons in the PNS PSR J0348+0432}
\begin{figure}[!htp]
\centering{}\includegraphics[width=4.5in]{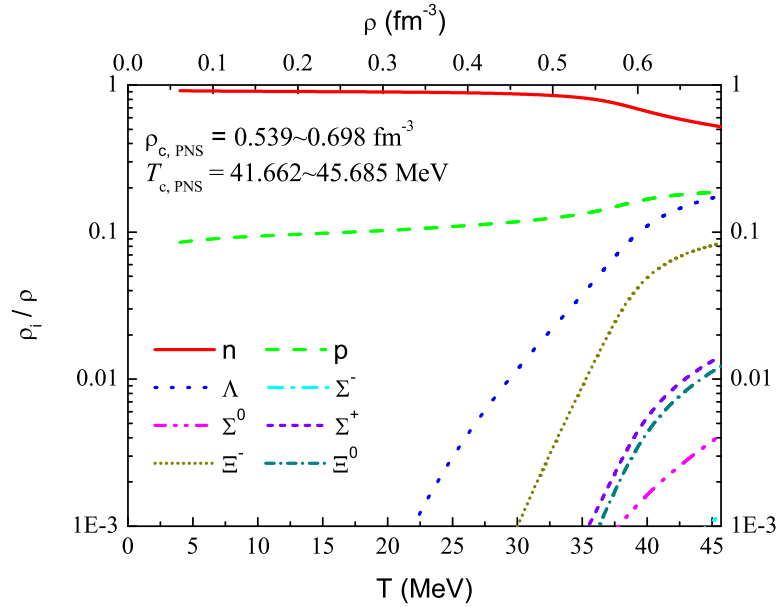}\caption{Effect of temperature on the relative particle number density of the baryons in the PNS PSR J0348+0432.}
\label{fig7}
\end{figure}

Effect of temperature on the relative particle number density of the baryons in the PNS PSR J0348+0432 is shown in Fig.~\ref{fig7}. We see that the central temperature of the PNS PSR J0348+0432 is in the range $T_{c, PNS}$=41.662$\sim$45.685 MeV corresponding to the relative particle number density of the baryons in the center $\rho_{c, PNS}$=0.539$\sim$0.698 fm$^{-3}$.

From Fig.~\ref{fig7} we also see that the relative particle number density of neutrons decreases whereas those of p, $\Lambda$, $\Sigma^{-}$, $\Sigma^{0}$, $\Sigma^{+}$, $\Xi^{-}$ and $\Xi^{0}$ all increase with the increase of the temperature.

\section{Conclusions}
In this paper, we apply the RMF theory to study the property difference between the NS PSR J0348+0432 and its PNS. The entropy per baryon is selected as $S$=2.

We find that the central baryon number density of the PNS PSR J0348+0432 should be in the range $\rho_{c,PNS}$=0.539$\sim$0.698 fm$^{-3}$, while that of the NS PSR J0348+0432 is in the wider range $\rho_{c,NS}=0.634\sim0.859$ fm$^{-3}$.

The field strengths of mesons $\sigma, \omega$ and $\rho$ in the PNS PSR J0348+0432 and those in the NS PSR J0348+0432 make little difference corresponding to a same baryon number density $\rho$. But the value range of the central field strength of mesons $\sigma, \omega$ and $\rho$ in the PNS PSR J0348+0432 is smaller than those in the NS PSR J0348+0432.

The central chemical potentials of neutrons and electrons in the PNS PSR J0348+0432 are all smaller than those in the NS PSR J0348+0432.

In our calculations baryon octet is considered. we find that inside the NS PSR J0348+0432 only neutrons, protons, $\Lambda$ and $\Xi^{-}$ produce. But in the PNS PSR J0348+0432, hyperons $\Sigma^{-}$, $\Sigma^{0}$, $\Sigma^{+}$ and $\Xi^{0}$ will also produce, though their relative particle number density are very small.

The higher temperature would be advantageous to the production of the hyperons $\Sigma^{-}$, $\Sigma^{0}$, $\Sigma^{+}$ and $\Xi^{0}$.

In this work, we consider the PNS as static star and the effects of rotation on PNS are not included.
But the research results show that the conservation of baryon number and angular momentum will determine the maximum frequencies of rotation during the cooling~\cite{Franzon16,Dexheimer08}. Therefore, in the next work we should study the effect of rotation on the properties of the PNS PSR J0348+0432.

\section*{Acknowledgments}
We are thankful to Shan-Gui Zhou for fruitful discussions during my visit to the Institute of Theoretical Physics of Chinese Academy of Sciences.
This work was supported by the Natural Science Foundation of China ( Grant No. 11447003 ) and the Scientific Research Foundation of the Higher Education Institutions of Anhui Province, China ( Grant No. KJ2014A182 ).

\bibliography{apssamp}

\end{document}